# Fractional occupation in Kohn-Sham density-functional theory and the treatment of non-pure-state *v*-representable densities.

Eli Kraisler

*Raymond and Beverley Sackler Faculty of Exact Sciences, School of Physics and Astronomy, Tel Aviv University, Tel Aviv 69978, Israel and Physics Department, NRCN, P.O. Box 9001, Beer Sheva 84190, Israel*

Guy Makov

*Physics Department, Kings College, London, The Strand, London WC2R 2LS, UK and Physics Department, NRCN, P.O. Box 9001, Beer Sheva 84190, Israel*

Nathan Argaman

*Physics Department, NRCN, P.O. Box 9001, Beer Sheva 84190, Israel*

Itzhak Kelson

*Raymond and Beverley Sackler Faculty of Exact Sciences, School of Physics and Astronomy, Tel Aviv University, Tel Aviv 69978, Israel*

In the framework of Kohn-Sham density-functional theory, systems with ground-state densities that are not pure-state *v*-representable in the non-interacting reference system (PSVR) occur frequently. In the present contribution, a new algorithm, which allows the solution of such systems, is proposed. It is shown that the use of densities which do not correspond to a ground state of their non-interacting reference system is forbidden. As a consequence, the proposed algorithm considers only non-interacting ensemble *v*-representable densities. The Fe atom, a well-known non-PSVR system, is used as an illustration. Finally, the problem is analyzed within finite temperature density-functional theory, where the physical significance of fractional occupations is exposed and the question of why degenerate states can be unequally occupied is resolved.

31.15.E-, 31.15.A-, 31.15.-p

# I. INTRODUCTION.

Density-Functional Theory (DFT) is the leading theoretical framework for studying the electronic properties of matter [1, 2, 3]. Most practical applications rely on the Kohn-Sham (KS) formulation [4], which was originally formulated assuming that the ground state density of an interacting $N$-electron system in an external potential $v(\vec{r})$ can be represented by the ground state density of a non-degenerate non-interacting $N$-electron system in a reference potential $v_{eff}(\vec{r})$. This assumption is known as pure-state $v$-representability of the density in the non-interacting system, and is denoted PSVR below.

For PSVR systems [4], the $N$-electron wave function $\Psi(\vec{r}_1,...,\vec{r}_N)$ of the reference system can be written as a single Slater determinant of the one-electron wave functions $\psi_i(\vec{r})$, which are solutions of the non-interacting Schrödinger equation $\left((-\hbar^2/2m_e)\nabla^2 + v_{eff}(\vec{r})\right)\psi_i = \varepsilon_i\psi_i$. The index $i$ refers to the relevant set of quantum numbers, e.g. $i = \{n,l,m,\sigma\}$ for spherical atoms. The density is given by $n(\vec{r}) = \sum_i^N |\psi_i(\vec{r})|^2$, where the sum is over the $N$ lowest-energy eigenstates of the system. The effective potential $v_{eff}[n] = v(\vec{r}) + v_H[n] + v_{xc}[n]$ is a functional of the density, where $v_H[n] = e^2 \int n(\vec{r}')/|\vec{r}-\vec{r}'| d^3r'$ and $v_{xc}[n] = \delta E_{xc}/\delta n$ are the Hartree and the exchange-correlation potentials, respectively. The density and the total energy $E[n] = T_s[n] + \int n(\vec{r})v(\vec{r})d^3r + E_H[n] + E_{xc}[n]$ of the interacting system are found by solving self-consistently the equations above. Here $T_s[n] = -\hbar^2/(2m_e)\sum_i^N \langle\psi_i|\nabla^2|\psi_i\rangle$ is the kinetic energy and $E_H[n]$ and $E_{xc}[n]$ are the Hartree and exchange-correlation energy terms. It was assumed, and later shown to be true [5, 6], that for ground state densities the exchange-correlation energy derivative $\delta E_{xc}/\delta n$ exists. Note, however, that the self-consistent solution of the Kohn-Sham equations requires the existence of the exchange-correlation energy derivative $\delta E_{xc}/\delta n$ for any density considered in the iterative process, not only for the ultimate density.

Two questions arise immediately from this formulation, the first of which has been considered extensively and the second far less so. (i) Are all ground state densities PSVR? (ii) For which densities does $v_{xc}[n] = \delta E_{xc}/\delta n$ exist and how can the self-consistent solution process ensure that only such densities are considered? From early on it was known [7, 8] that there are physical systems (e.g., Fe and

Co atoms in the spherical approximation), for which the density was not represented by a pure state. For these atoms, no self-consistent result was found for which integrally occupied orbitals were also the lowest in energy. The absence of a self-consistent solution expressed itself in the so-called 'Fermi statistics' problem [9]: the energy levels of the effective potential cross repeatedly during the iterations, and being occupied according to the ground-state rule, yield radically different orbital densities, so that the self-consistent cycle cannot converge.

A possible stopgap measure is to suspend the requirement of ground state occupation and to allow the system to relax while freezing the energy level occupation, resulting in convergence to a solution in which the representing non-interacting system is in an excited state. In this spirit, Janak [8] proposed an extension of the Kohn-Sham scheme to include occupation numbers $\{g_i\}$ (termed the electronic configuration) as parameters in the definitions of the density $n(\vec{r}) = \sum_i g_i |\psi_i(\vec{r})|^2$ and the kinetic energy $\tilde{T}_s = \sum_i g_i \langle \psi_i | -\frac{\hbar^2}{2m_e} \nabla^2 | \psi_i \rangle$. To determine the occupation numbers, Janak minimized the total energy $\tilde{E}[\{g_i\}] = \tilde{T}_s + \int n(\vec{r}) v(\vec{r}) \, d^3r + E_H[n] + E_{xc}[n]$ with respect to $\{g_i\}$. It was shown [8] that the minimum of the total energy $\tilde{E}$ is achieved only for a "proper" electronic configuration, i.e. one in which the energy levels are occupied without gaps. This includes not only the PSVR systems treated by Kohn and Sham, but also cases where the representing non-interacting $N$-electron system possesses a ground-state degeneracy, with the highest electron orbitals degenerate and partially occupied. For such a configuration the density is ensemble $v$-representable in the non-interacting reference system [2], and this is denoted by NI-EVR below. However, Janak questioned the validity of his results in the non-PSVR cases, especially in situations with unequal occupations of the degenerate orbitals.

Addressing the problem of $v$-representability, a recent study by Ullrich and Kohn [10] on topologies of the $v$ and $n$ spaces examined for lattice systems showed that potentials which generate NI-EVR densities are not isolated points in $v$-space, because not every perturbation potential imposed on the system will lift the degeneracy of its energy levels. Even more significant is that NI-EVR densities occupy manifolds of finite volume in $n$-space. Thus, systems with NI-EVR densities are not rare, but form a significant physical class that should be treated with proper care.

Few examples of EVR densities have been studied in detail. Some of these studies are analytical, and thus relate to the exact (but unknown) exchange-correlation functional, while others involve direct computations using available approximate functionals. Recently, two analytical examples were introduced [10, 11]: a finite lattice problem, which produced either a PSVR or an NI-EVR solution, depending on the value of the potential on the different lattice points, and the Be ion series.

Upon application of the Janak formalism, NI-EVR densities have been identified in several physical systems, such as Fe and Co atoms [7, 8]. The occupations of the $3d^\downarrow$ and $4s^\downarrow$ degenerate levels are found to be fractional and unequal, contrary to what one might expect from elementary statistical mechanics considerations. This surprising result made the authors [7, 8] question their findings and speculate that they might be a consequence of the spherical symmetry approximation or the local-density approximation employed.

Other authors have rejected fractional occupation altogether. Averill and Painter [12] devised a numerical scheme in which the minimization over the occupations in Janak's functional is performed as part of the iterative process, but took the view that the physical representation of the system would necessarily involve an electronic configuration with integral occupation numbers, even if it is "improper". Alternatively, in a well-known comprehensive study in atomic physics performed by Kotochigova et al. [13], the experimental electronic configurations were used throughout the calculations. Although for some atoms the occupations thus obtained are "improper", the validity of such a choice of electronic configuration for different atoms/ions was not discussed.

The Kohn-Sham scheme requires differentiability of both the kinetic and the exchange-correlation functionals [5, 6]. After the generalization of density-functional theory in the constrained-search formalism to all *N*-representable densities [14, 15], it was found [5, 6, 16] that differentiability exists only for densities which are both interacting ensemble *v*-representable (I-EVR) and non-interacting ensemble *v*-representable (NI-EVR) (see Sec. II for the definitions). However, during the iterative convergence of the Janak-Kohn-Sham equations, the densities generated are not restricted to be EVR. This is unlike the situation in the self-consistent solution of the Kohn-Sham equations, where if the densities are generated from ground states of the trial effective potentials, then they are NI-EVR by construction. Therefore, the Janak scheme is expected to be invalid for the exact form of $E_{xc}[n]$ and is formally applicable only for approximations, which are differentiable at all densities, such as the local-density approximation (LDA).

Therefore, in consequence of the extension of the DFT-KS formalism to include NI-EVR densities, it is desirable to have an algorithm which will guarantee that only EVR densities are considered in the self-consistent solution of the Kohn-Sham equations. To the best of our knowledge, such an algorithm has not yet been suggested. We note that it was proved [5, 17] that for lattice systems all the densities are EVR. Therefore, in computational implementations on a grid all densities are implicitly EVR. However, this result has not been extended to the continuum, and this indicates the benefit of an algorithm which is restricted to explicitly EVR densities.

Surprisingly, in all of these developments the finite-temperature version of DFT (T-DFT) is scarcely mentioned. The finite-temperature theory can be represented as an extension of the standard theory of thermal ensembles to non-uniform systems [3]. It is well-known that finite-temperature DFT does not suffer from the *v*-representability problem, and connects continuously to the ground-state theory in the limit $T \to 0$ [18]. This has previously allowed the resolution or reinterpretation of several theoretical issues in DFT.

In the present work we present three contributions to the ongoing discussion. First, we show that densities generated by "improperly" occupying the orbitals in the reference system cannot be ground state densities of the given external potential, complementing the known result that the ground state can be described by properly occupying the reference system. Consequently, we present a new algorithm which allows the solution of Kohn-Sham systems with ground-state densities that are not PSVR, while considering only EVR densities in the process and overcoming the "Fermi statistics" convergence difficulties [9]. We apply the algorithm to the well-known case of Fe and show that the results obtained are converged, stable numerically and the emergence of NI-EVR density is insensitive to the choice of the exchange-correlation functional approximation. We also demonstrate that the subspace of external potentials which yield NI-EVR densities is finite, by considering variations of the nuclear charge about $Z = 26$. In passing, we discuss the solution of the Be ion series in the LDA. Finally, our third contribution is an analysis of the EVR problem within T-DFT, where we expose the physical significance of the fractional occupations and resolve the question of why degenerate states can be unequally occupied.

## II. THEORY.

The extension of the Kohn-Sham formalism to EVR densities is briefly presented below for completeness. An interacting ensemble-*v*-representable (I-EVR) density is defined [5] as an integrable non-negative function $n(\vec{r}) = \sum_k \lambda_k n^{(k)}(\vec{r})$, with $\lambda_k \geq 0$, $\sum_k \lambda_k = 1$, where

$n^{(k)}(\vec{r}) = N \int ... \int |\Psi^{(k)}(\vec{r}, \vec{r}_2, ..., \vec{r}_N)|^2 d\vec{r}_2 ... d\vec{r}_N$ and $\Psi^{(k)}$ are orthonormal degenerate ground states of the Hamiltonian of an interacting many-electron system :

$$\hat{H} = \hat{T} + \hat{V} + \hat{W} \qquad (1)$$

Here $\hat{T} = -\hbar^2/(2m_e)\sum_i \nabla_i^2$ is the kinetic energy operator, $\hat{V} = \sum_i v(\vec{r}_i)$ is the external potential operator and $\hat{W} = (e^2/2)\sum_{i \neq j} 1/|\vec{r}_i - \vec{r}_j|$ is the electron-electron repulsion operator. The pure-state case is a particular case with only one $\lambda_k = 1$, and the others are zero. A non-interacting ensemble-v-representable (NI-EVR) density is defined similarly, with a non-interacting Hamiltonian $\hat{H}_0 = \hat{T} + \hat{V}_{eff}$ ($\hat{W} = 0$).

According to the constrained search formulation of Hohenberg-Kohn theorems, the ground state energy is the minimum value of the functional of the density $E_v[n] = \int v(\vec{r}) n(\vec{r}) d^3r + F_L[n]$, for a given external potential $v(\vec{r})$. $F_L[n]$ is the Lieb functional ([2], p. 14 and references therein, [13], [16]):

$$F_L[n] = \min_{\sum_i d_i n_i = n} \left[ \sum_i d_i \langle \Phi_i | \hat{T} + \hat{W} | \Phi_i \rangle \right]. \qquad (2)$$

with $d_i \geq 0$, $\sum_i d_i = 1$, $n_i(\vec{r}) = N \int ... \int |\Phi_i(\vec{r}, \vec{r}_2, ..., \vec{r}_N)|^2 d\vec{r}_2 ... d\vec{r}_N$, where $\Phi_i(\vec{r}_1, \vec{r}_2, ..., \vec{r}_N)$ are all anti-symmetric normalized N-particle wave functions.

Similarly to the pure-state case, the density is represented through a reference system of non-interacting electrons, defined by $v_{eff}(\vec{r})$. Then, the energy $E_v[n]$ can be written as $E_v[n] = \int v(\vec{r}) n(\vec{r}) d^3r + T_L[n] + E_H[n] + E_{xc,L}[n]$, where

$$T_L[n] = \min_{\sum_i d_i n_i = n} \left[ \sum_i d_i \langle \Phi_i | \hat{T} | \Phi_i \rangle \right], \qquad (3)$$

$E_H[n]$ is the Hartree energy term and the exchange-correlation term is defined as $E_{xc,L}[n] = F_L[n] - E_H[n] - T_L[n]$.

To obtain the density from the reference system, the effective potential is determined, up to a constant, via the differential Euler relations for the interacting and the non-interacting systems :

$$\delta T_L/\delta n + v(\vec{r}) + v_H[n] + \delta E_{xc,L}/\delta n = const, \qquad (4)$$

$$\delta T_L/\delta n + v_{eff}(\vec{r}) = const.$$

It follows that $v_{eff}(\vec{r}) = v(\vec{r}) + v_H[n] + v_{xc,L}[n]$, only if both $\delta T_L/\delta n$ and $\delta E_{xc,L}/\delta n = v_{xc,L}[n]$ exist. Otherwise, no relation between the interacting system and the reference system can be established.

The differentiability conditions for both $F_L[n]$ and $T_L[n]$ were formulated by Englisch and Englisch [5]: (1) the functional $T_L[n]$ is differentiable at all NI-EVR densities, and only there; (2) the functional $E_{xc,L}[n]$ is differentiable at densities, which are both NI-EVR and I-EVR, and only there. While the equivalence of these two sets of densities has the status of a conjecture, it can be proved that for any I-EVR density there exists a NI-EVR density arbitrarily close. Therefore, a Kohn-Sham scheme can always be set up [16]. Regarding condition (2), note that for explicit approximations employed for $E_{xc,L}[n]$, such as the LDA, mathematical differentiability is always assured, even if for the given density the exact functional would not be differentiable.

As to condition (1), we note that the ground state wave-functions $\Phi_i$ of a non-interacting system must obviously be constructed as Slater determinants of the $N$ lowest eigenfunctions $\psi_i(\vec{r})$ that emerge from the Schrödinger equation $\left(-(\hbar^2/2m_e)\nabla^2 + v_{eff}(\vec{r})\right)\psi_i = \varepsilon_i\psi_i$. It can be shown that the kinetic energy functional $T_L[n]$ takes the form

$$T_L[n] = \min_{\sum_i g_i |\psi_i|^2 = n} \left[ \sum_i g_i \left\langle \psi_i \left| -\frac{\hbar^2}{2m_e}\nabla^2 \right| \psi_i \right\rangle \right] \qquad (5)$$

and the density is given by

$$n(\vec{r}) = \sum_i g_i |\psi_i(\vec{r})|^2 , \qquad (6)$$

where the occupation numbers $g_i$ are linearly related to the coefficients $d_i$ and are restricted to be :

$$g_i = \begin{cases} D_i & : \quad \varepsilon_i < \varepsilon_F \\ x_i & : \quad \varepsilon_i = \varepsilon_F \\ 0 & : \quad \varepsilon_i > \varepsilon_F \end{cases} . \qquad (7)$$

Here $x_i \in [0, D_i]$, $\varepsilon_F$ is the highest occupied energy level and $D_i$ is the degeneracy of the $i$-th level, which can now be greater than 1, depending on the special symmetries the problem may have, e.g. spherical symmetry for atoms. The restriction (7) is a straightforward consequence of the fact that only ground states of a reference system are considered. We denote occupations consistent with (7) as "proper".

It can be seen from (7) that fractional occupation can occur only if for more than one energy level $\varepsilon_i = \varepsilon_F$. The physical interpretation of such a situation is discussed by us in Sec. IV.

Now we consider the use of "improper" densities, i.e. those which are excited states of their own effective potential. As frequently happens in Janak-type calculations [7, 8, 12], a trial density $n(\vec{r})$ is represented in the form (6) with an "improper" (excited) occupation of levels of a non-interacting system, defined by $v_{eff}(\vec{r})$, which obeys $v_{eff}(\vec{r}) = v(\vec{r}) + v_H[n] + v_{xc}[n]$. It is not evident that in such a case the derivative $\delta T_L / \delta n$ must exist [5]. However, if it exists, then there exists another non-interacting system defined by $\tilde{v}_{eff}[n] = \tilde{v}(\vec{r}) + v_H[n] + v_{xc}[n]$, for which $n(\vec{r})$ is the ground-state density or a linear combination of ground-state densities. Thus, $n(\vec{r})$ can be represented in terms of the (˜)-system as $n = \sum_i \tilde{g}_i |\tilde{\psi}_i|^2$ with "proper" occupation. Then, however, $n(\vec{r})$ cannot be the ground state of the original system of interest because Euler's relations (4) cannot be satisfied for $v_{eff}[n]$ if they were satisfied for $\tilde{v}_{eff}[n]$. Therefore, we arrive at the conclusion that an "improper" configuration cannot be used in (6) to construct the density, even if $\delta T_L / \delta n$ exists, as could be misunderstood from Ref. 19. As a corollary, an algorithm that relies on the generalized form $T_L[n]$ must be restricted to "proper" densities only.

Considering the validity of Janak's theorem questioned by Valiev and Fernando [19], we comment that Janak's theorem is valid, but only for Janak's $\tilde{E}$, which is an extension of the energy functional to include systems with variable and fractional number of electrons, which are beyond the scope of the present work. In the domain of systems with integer occupation, the ground state densities are restricted to be "proper", as in (7) and the $\{g_i\}$ are determined uniquely by the self-consistency requirement. In this formalism, Janak's theorem, in the sense of a derivative of the energy functional with respect to occupation numbers, is inapplicable and this, we believe, should be the context of Ref. 19.

The above formalism can be extended [2] to spin-polarized cases via representing a spin-polarized system as two distinct (but coupled) electronic subsystems, denoted by the index $\sigma = \{\uparrow, \downarrow\}$. The generalized kinetic energy functional $T_L[n_\uparrow, n_\downarrow] = T_L[n_\uparrow, 0] + T_L[n_\downarrow, 0]$, where

$$T_L[n_\sigma, 0] = \min_{\sum_i g_{i\sigma}|\psi_{i\sigma}|^2 = n_\sigma} \sum_i g_{i\sigma} \langle \psi_{i\sigma} | -\frac{\hbar^2}{2m_e}\nabla^2 | \psi_{i\sigma} \rangle.$$

Each of the subsystems is restricted to be "properly" occupied by (7), but there can be gaps in the occupation of the whole system if the first vacant level of one of the sub-systems lies below the last occupied level of the another one. We call such a system "proper in a broad sense". This situation arises because we are constrained from having partial occupation of spin states by the physical requirement that the total spin in the reference system is half-integer, i.e. that the spin occupations remain integral.

To summarize, the only densities which can be considered as candidate densities in the constrained search formalism are those with "proper" sets of occupation numbers. In the spin-polarized case, each of the spin subsystems must be properly occupied, even though the combined system may be "improperly" occupied due to the constraint of integral spin occupations. It remains to be shown in Sec. III how to restrict the self-consistent search for the density to NI-EVR densities only.

## III. NUMERICAL ALGORITHM AND APPLICATIONS.

In this section we present an algorithm to solve the Kohn-Sham equations that is restricted to searching only over explicitly NI-EVR densities in accordance with the results of Sec. II. We have shown above that Janak's approach involves minimizing the energy over a set of densities which are not necessarily "proper", and therefore the existence of $v_{xc}[n] = \delta E_{xc}/\delta n$ is not assured. We apply our algorithm to the spherical Fe atom which is a simple, well-established [7, 8, 12] non-PSVR system, which is NI-EVR. As an example of a NI-EVR system, we investigate some of its properties. In passing we show that the Be ion series which has been shown to be a NI-EVR system [10, 11, 20] is instead found to be PSVR in the LDA.

In the case of the spherical Fe atom, the use of integral occupation numbers does not lead to a "proper" solution of the Kohn-Sham equations, a fact that was troubling from early days [7, 8]. Allowing fractional occupation numbers determined by the minimization of the total energy in Janak's approach yields a NI-EVR density for the spin $S = 2$. E.g., using the Vosko, Wilk and Nusair (VWN) [21] parameterization of the LSDA for the exchange-correlation functional and a spherical approximation for the density yields $E = -1261.229308$ hartree as the total energy for Fe, with an electronic configuration of $[Ar]3d^5_{1.398}4s^1_{0.602}$ [22]. This result is in agreement with earlier publications on Fe [7, 8, 12].

Our alternative algorithm for treating NI-EVR densities is to start with a reasonable guess for $v_{eff,\sigma}(\vec{r})$ and occupy the energy levels of the reference systems "properly" given the total spin value $S$. Therefore, we are assured that the density generated by the reference system in each iteration is NI-EVR and by the EVR conjecture [16] is also I-EVR. This allows us to calculate the effective potential from this density. If we employ this potential directly to calculate a new density, then as expected [9], the process starts alternating between two electronic configurations, e.g. for $S = 2$ between $[Fe] = [Ar]3d_1^5 4s_1^1$ and $[Fe] = [Ar]3d_2^5 4s_0^1$.

Instead, as the first new step in the algorithm, we reduce the linear mixing coefficient for the effective potential each time the electronic configuration changes. The reduction factor can vary around 0.5. By repeating this procedure, the eigenvalues eventually stop crossing and slowly approach each other. It is possible then to find an effective potential, with energy levels ($\varepsilon_{3d\downarrow}$ and $\varepsilon_{4s\downarrow}$ in this example) that coincide, to any required accuracy ($10^{-6}$ hartree was required for Fe) with the original integer configuration. However, this solution is not necessarily a self-consistent one, and additional iterations in the domain of NI-EVR densities are required to obtain the final result.

The second new step in the algorithm continues the search in the subspace of potentials that generate NI-EVR densities while relaxing the constraint on integer occupation. Once an effective potential with two degenerate levels is found, it can produce a range of legitimate, "proper" NI-EVR densities by fractionally occupying the degenerate levels. Utilizing the fact that potentials generating NI-EVR densities are not, in general, points in $v_{eff}$-space, we choose the occupation that will generate a new potential which preserves the degeneracy of the energy levels for the next iteration. Assuming the potential change between two consecutive iterations (denoted $k$ and $k+1$) $w = v_{eff}^{(k+1)}[n^{(k+1)}(\{g_i\})] - v_{eff}^{(k)}$ is small, it can be shown by means of perturbation theory [23] that the degeneracy of levels $a$ and $b$ will be preserved if

$$\langle \psi_a^{(k)} | w | \psi_a^{(k)} \rangle - \langle \psi_b^{(k)} | w | \psi_b^{(k)} \rangle = 0. \tag{8}$$

Here $v_{eff}^{(k)}$, $n^{(k)}$ and $\psi_m^{(k)}$ are the effective potential, the density and the wavefunction of the $m$-th level, in the $k$-th iteration, respectively, and $\{g_i\}$ symbolizes the dependence of the density on fractional occupations. Note that the presence of a Fermi-statistics problem implies that equality (8) is fulfilled within the available range, as the left-hand side changes sign as a function of the occupations, with the degeneracy of the levels $a$ and $b$ lifted in one direction for one extreme occupation, and in the other direction for the other extreme. Numerically, equality (8) is enforced by searching for the electronic configuration $\{g_i\}$ that, once employed to obtain a new density $n^{(k+1)}(\{g_i\})$, and consequently $v_{eff}^{(k+1)}[n^{(k+1)}(\{g_i\})]$, will nullify the left-hand side of eq. (8). The search is made using Newton-Raphson's method, which converges within 3-5 steps for Fe, where we have required a convergence of $10^{-6}$ for the occupation numbers. This search is repeated for each iteration in this part of the convergence process, affording us to preserve the degeneracy in the reference system.

Continuing the iterations described above, while preserving the degeneracy of the energy levels ($\varepsilon_{3d\downarrow}$ and $\varepsilon_{4s\downarrow}$ in Fe), but not constraining their value, we eventually converge to a self-consistent result, which is confirmed by increasing the mixing coefficient. This result for Fe coincides, within the convergence accuracy of $\Delta_E = 10^{-6}$ hartree, with the result reached by Janak's approach.

A schematic description of convergence of Janak's algorithm and the algorithm proposed above in the $v_{eff} - n$ plane is shown in Fig. 1. Janak's algorithm passes through "improper" densities. A full minimization of occupation numbers brings it, however, to a "proper" fractionally occupied configuration. The algorithm proposed here uses only "proper" configurations, and by reduction of the mixing constant, enters the NI-EVR range of potentials, where it converges to the desired density.

The process above is performed for all values of the total spin *S*. In the example of Fe, the energy levels for $S = 0,1$ are occupied "properly in a broad sense" and possess higher total energies, as shown in Table I. As mentioned above, the stability of the results obtained was verified by increasing the mixing coefficient once the iterations converged. No separation of the degenerate energy levels was observed. The influence of the initial guess for the effective potential on the final result was also checked. It was found that the eigenvalue at which the energy levels first became degenerate depends significantly on the guess. However, during the subsequent iterative process within the degenerate subspace, the eigenvalues converge to the expected value, along with the total energy and other quantities. Therefore, we conclude that the initial condition affects only the number of iterations needed to reach self-consistency, and not the converged self-consistent values.

In order to demonstrate that an emergence of a NI-EVR solution for Fe is not an artifact related to the local-density approximation for the exchange-correlation functional as speculated in Refs. 7 and 8, we repeated the calculation adding the PBE-GGA [24] to the VWN-LSDA. We obtained a NI-EVR density in this approximation as well, with $E = -1263.446126$ hartree for the total energy and $[Ar]3d^5_{1.349}4s^1_{0.651}$ for the electronic configuration, as can be seen from Table I. Moreover, we have applied our algorithm successfully to additional atoms and ions using the same procedures described here. The details of these calculations are omitted for brevity.

It was emphasized recently [10] that NI-EVR densities are not points in $v$-space, i.e. given an external potential $v(\vec{r})$ that produces a NI-EVR density, there are potentials in its environment that also produce NI-EVR densities. We use the Fe atom to demonstrate such a situation. Changing the nucleus' charge $Z$, we obtain a region $25.82 < Z < 26.25$ for which NI-EVR densities occur, whereas for $Z < 25.82$ or $Z > 26.25$ PSVR solutions are obtained. The energy levels of the reference system and the occupation numbers are plotted in Fig. 2 and 3, as a function of $Z$.

Finally, we comment on the Be ion series that has been previously presented as an example of a NI-EVR system. It was proposed [10, 11] to solve the Be ion series in the limit of large $Z$ analytically, by mapping it onto a four-electron system with an electron-electron interaction, which becomes negligible in the limit $Z \to \infty$. The mapping is performed by a scaling $\tilde{r} = Zr/4$ for the distances and $\tilde{e}^2 = 4e^2/Z$ for the electron charge, in the Hamiltonian (1). When $Z \to \infty$, the electron-electron interaction in the $(\sim)$-system vanishes and the energy levels $2s$ and $2p$ become degenerate, with a first-order correction proportional to $1/Z$. If the perturbation applied is of the exact form $e^2/|\vec{r}_1 - \vec{r}_2|$, then the correct zeroth order wave functions mix the $(2s)^2$ and $(2p)^2$ configurations and generate the NI-EVR density as reported by Ullrich and Kohn [10] and in agreement with the CI-based procedure of Morrison [20]. However, Kohn-Sham calculations do not support this result, for reasons unknown [20]. We propose that this disagreement arises from symmetry considerations. If a Hartree type approximation modified by the LDA (or an extension) is the perturbation, then its spherical character will cause the relevant wave functions to separate into the $(2s)^2$ and $(2p)^2$ configurations, thus leading to a PSVR type density. Such a separation can be seen in Fig. 4, which illustrates the difference between the energy levels $\varepsilon_{2s}$ and $\varepsilon_{2p}$ of the Be ion series calculated for $Z = 4, 40, 100, 200$ with VWN-LSDA.

## IV. DISCUSSION. T-DFT ANALYSIS.

A better understanding of ensemble *v*-representable systems can be achieved by considering the finite-temperature approach to DFT [3, 10, 25], in which ground-state DFT corresponds to the limit of vanishing temperature [18].

It can be shown (see [3] and references therein) that at finite temperatures the grand canonical potential

$$\Omega[v(\vec{r})] = -k_B T \ln\left[\text{Tr}\left(\exp\left(-\frac{\hat{T}+\hat{W}+\hat{\rho}v}{k_B T}\right)\right)\right] \quad (9)$$

is convex with respect to $v(\vec{r})$ and is always differentiable (for definitions see Sec. II). Here $\hat{\rho} = \sum_{i=1}^{N}\delta(\vec{r}-\vec{r}_i)$ is the density operator, whose statistical average yields the electron density $n(\vec{r}) = \langle\hat{\rho}\rangle$. The first functional derivative of $\Omega$ equals the density: $\delta\Omega/\delta v(\vec{r}) = n(\vec{r})$. As a result, the Hohenberg-Kohn free-energy functional $F_{HK}[n(\vec{r})] = \Omega[v(\vec{r})] - \int n(\vec{r})v(\vec{r})d^3r$ is differentiable with respect to the density, and its first derivative is $\delta F_{HK}/\delta n(\vec{r}) = -v(\vec{r})$. Since for a non-interacting electron system, $F_{HK}[n(\vec{r})]$ reduces to the kinetic energy functional, this last functional is differentiable as well. Therefore, for electronic systems at finite temperatures the *v*-representability question does not exist.

In T-DFT, a many-electron interacting system at a temperature *T* is described by reference to a non-interacting system with the same temperature [26]. It follows then that the occupation numbers in the reference system are determined by the Fermi-Dirac distribution and the entropy is given by $S = -k_B \sum_i D_i\left(f_i \ln(f_i) + (1-f_i)\ln(1-f_i)\right)$, where the sum is over all energy levels and $f_i = g_i/D_i$. The effective potential preserves the form $v_{eff}(\vec{r}) = v(\vec{r}) + v_H[n] + v_{xc}[n]$, but now the exchange-correlation potential depends also on *T*.

At finite temperature the occupancy $g_i = D_i(1 + \exp[(\varepsilon_i - \mu)/k_B T])^{-1}$ is determined by the combination $(\varepsilon_i - \mu)/k_B T$, which would lead us to expect that the fractional occupations of degenerate states will be equal. However, closer observation shows that in the limit of zero temperature the Fermi-Dirac distribution becomes multi-valued for $\varepsilon_i = \mu$, taking on all values $0 \le g_i \le D_i$.

For cases with several degenerate states at the Fermi level having different occupations, the degeneracy is lifted linearly with temperature, with the combinations $(\varepsilon_i - \mu)/k_B T$ left essentially unchanged. This can be seen in detail from the following relations. Expanding the energy and the chemical potential for low temperatures, we obtain $\varepsilon_i(T) = \varepsilon_i^{(0)} + \varepsilon_i^{(1)}T + \frac{1}{2}\varepsilon_i^{(2)}T^2$, $\mu(T) = \mu^{(0)} + \mu^{(1)}T + \frac{1}{2}\mu^{(2)}T^2$ up to terms $O(T^3)$, where $\varepsilon_i^{(n)} = (d^n\varepsilon_i/dT^n)_{T=0}$ and $\mu^{(n)} = (d^n\mu/dT^n)_{T=0}$. If at $T=0$ there are two or more degenerate partially-occupied levels, whose energy is $\varepsilon^{(0)}$, then obviously $\mu^{(0)} = \varepsilon^{(0)}$. Then, the partial occupation numbers of the $i$-th level at zero temperature are determined by the relation:

$$g_i^{(0)} = D_i \lim_{T \to 0}\left[(1+\exp[(\varepsilon_i - \mu)/k_B T])^{-1}\right] = D_i(1+\exp[(\varepsilon_i^{(1)} - \mu^{(1)})/k_B])^{-1}. \quad (10)$$

This exposes the physical significance of the magnitude of the partial occupation as indicating the temperature dependence of the eigenstate relative to that of the chemical potential. States with occupation less than $D_i/2$ will be more temperature-dependent than the chemical potential and vice versa. Only states which are both degenerate (i.e. have the same value of $\varepsilon_i^{(0)}$) and physically equivalent (i.e. have the same value of $\varepsilon_i^{(1)}$) will be equally occupied, whereas if the states are accidentally degenerate then the occupation numbers will be unequal, thus resolving Janak's conundrum [8]. Expanding the occupation numbers $g_i$, we obtain the following relation:

$$g_i = g_i^{(0)} - \frac{1}{2}g_i^{(0)}\left(1 - \frac{g_i^{(0)}}{D_i}\right)\left(\varepsilon_i^{(2)} - \mu^{(2)}\right)T + ... \quad (11)$$

It can be seen that the occupation depends linearly on the temperature through the second derivatives of the energy and the chemical potential, therefore varying significantly only at high temperatures.

We have conducted some finite-temperature calculations for Fe, which support the above explanation. We approximated the temperature-dependent exchange-correlation functional by its zero-temperature limit $F_{xc}[n](T) = E_{xc}[n] + O(T^2)$ [25], and used the PBE-GGA approximation, as above. We concentrated on the $S=2$ case only. Figure 5 shows the dependence of the eigenvalues $\varepsilon_{3d\downarrow}$ and $\varepsilon_{4s\downarrow}$ and the chemical potential as a function of the temperature. Figure 6 demonstrates the dependence of the normalized occupation numbers $f_i$ on the temperature. The numerical results agree with the expansions above. In particular, it can be seen that for a vanishing temperature the results for $\varepsilon_{3d\downarrow}^{(0)} = \varepsilon_{4s\downarrow}^{(0)} = \mu_\downarrow^{(0)} = -0.154624$ hartree, $g_{3d\downarrow} = 5f_{3d\downarrow} = 1.349$ and $g_{4s\downarrow} = f_{4s\downarrow} = 0.651$ reproduce the zero-temperature DFT results presented in Sec. III. In addition, from fig. 5 we deduce $\varepsilon_{3d\downarrow}^{(1)}/k_B = 3.0869$ hartree/K, $\varepsilon_{4s\downarrow}^{(1)}/k_B = 1.4659$ hartree/K and $\mu_\downarrow^{(1)}/k_B = 2.0909$ hartree/K. Substitution into equation (10) yields $g_{3d\downarrow}^{(0)} = 1.349$ and $g_{4s\downarrow}^{(0)} = 0.651$, in full agreement with the results from Sec. III.

## V. CONCLUSIONS.

In conclusion, in the present contribution we introduce a new algorithm for the self-consistent solution of the Kohn-Sham equations in cases of non-PSVR densities. Since we have shown that densities which are constructed from excited states of the effective potential cannot be ground state densities of the interacting system, our algorithm solves for non-PSVR systems while considering NI-EVR densities only. It is perhaps surprising that such algorithms have hitherto scarcely been discussed. Avenues for future research include further development of such algorithms, e.g., based on studies of the efficiency of their performance.

The Fe atom was used as an illustration. By considering variations in the nuclear charge $Z$ in Fe, we have also provided a demonstration that the subspace of external potentials which yields NI-EVR densities in Fe is finite. In addition, the fact that NI-EVR densities appear in Fe was found to be insensitive to the particular approximation of the exchange-correlation functional.

Lastly, the physical meaning of fractional occupation numbers, and in particular the unequal occupation of degenerate levels, has been clarified through the finite-temperature density-functional formalism. As the occurrence of ensembles in this formalism is natural, it may be expected that it will aid progress on additional issues as well, including the abovementioned issue of seeking improved algorithms.

## ACKNOWLEDGEMENT.

One of us (E.K.) acknowledges fruitful discussions with Mr. Nir Sapir.

| Configuration | S | XC | $\varepsilon(3d^\uparrow)$ (hartree) | $\varepsilon(3d^\downarrow)$ (hartree) | $\varepsilon(4s^\uparrow)$ (hartree) | $\varepsilon(4s^\downarrow)$ (hartree) | E (hartree) | "proper" |
|---|---|---|---|---|---|---|---|---|
| $[Ar]3d^5_{1.397}4s^1_{0.603}$ | 2 | LSDA | -0.275997 | -0.161152 | -0.196220 | -0.161151 | -1261.229308 | yes |
| $[Ar]3d^5_2 4s^0_1$ | 1 | LSDA | -0.187671 | -0.107214 | -0.155805 | -0.162969 | -1261.197603 | b. s. |
| $[Ar]3d^3_4 4s^1_0$ | 0 | LSDA | -0.140353 | -0.163057 | -0.168303 | -0.147304 | -1261.145890 | b. s. |
| $[Ar]3d^5_{1.349}4s^1_{0.651}$ | 2 | GGA | -0.277168 | -0.154624 | -0.191431 | -0.154623 | -1263.446126 | yes |
| $[Ar]3d^5_2 4s^0_1$ | 1 | GGA | -0.181669 | -0.096826 | -0.147664 | -0.158220 | -1263.411856 | b. s. |
| $[Ar]3d^3_4 4s^1_0$ | 0 | GGA | -0.131012 | -0.154871 | -0.163296 | -0.137957 | -1263.357365 | b. s. |

**TABLE I: The electronic configuration, energy levels and the total energy for Fe at different spin values solved within LSDA and GGA (see definitions in text). The cases are identified as "proper" or "proper in a broad sense" (b.s.).**

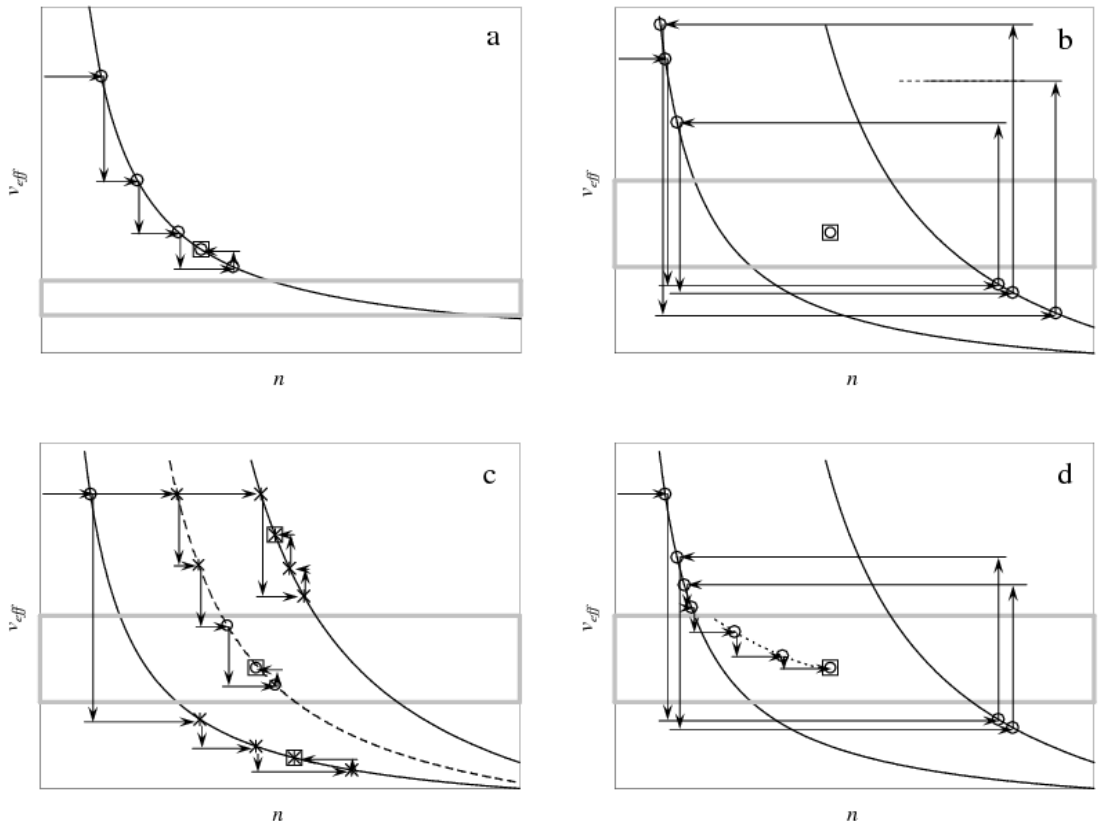

**FIG. 1.** Schematic representation of the numerical convergence with various algorithms in the $v_{eff}$ – $n$ plane. Each solid curve represents an integral electronic configuration. The dashed curve on panel (c) corresponds to a fractional electronic configuration. "Proper" occupations are denoted by circles, "improper"- by asterisks. Converged results ("proper" and "improper") are marked by squares. The region of NI-EVR effective potentials is denoted by a gray frame.

(a) A PSVR example. (b) A NI-EVR example. No convergence due to the 'Fermi statistics' problem. (c) A NI-EVR example. Janak's algorithm. (d) A NI-EVR example. The proposed algorithm.

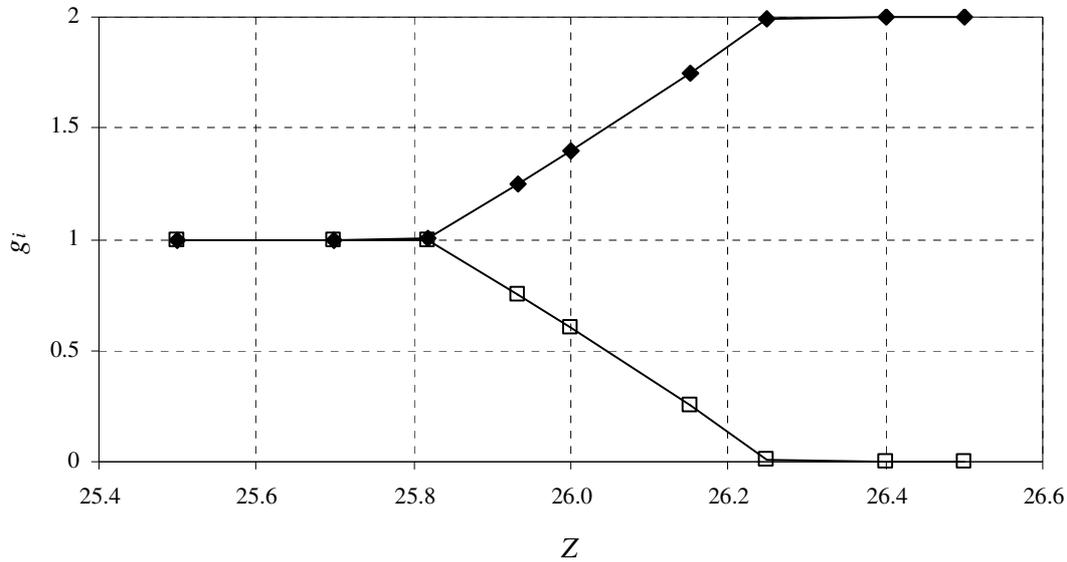

**FIG. 2.** The occupations of the $3d^{\downarrow}$ (rhombi) and $4s^{\downarrow}$ (squares) energy levels as a function of Z.

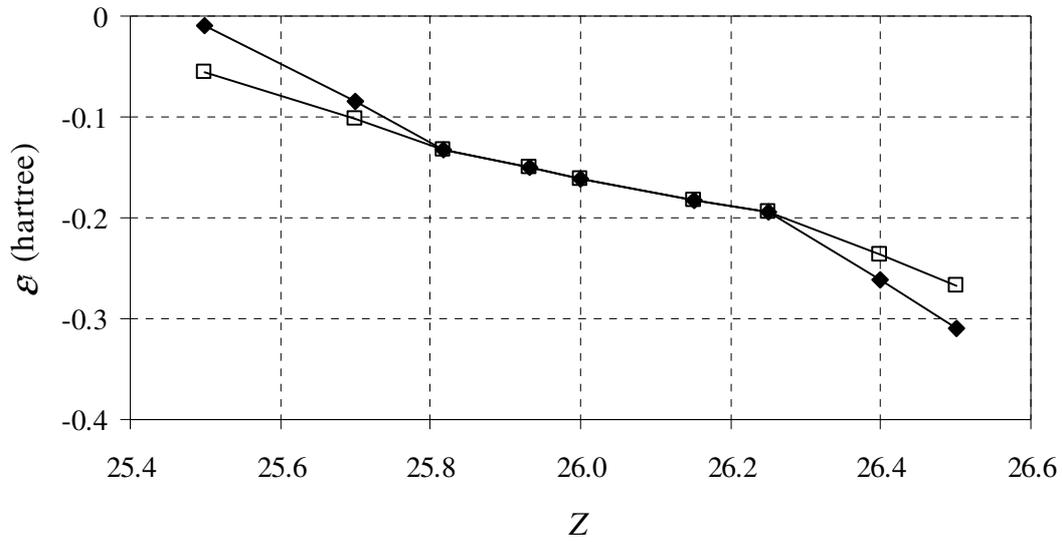

**FIG. 3.** The energy levels $\varepsilon_{3d\downarrow}$ (rhombi) and $\varepsilon_{4s\downarrow}$ (squares) as a function of Z.

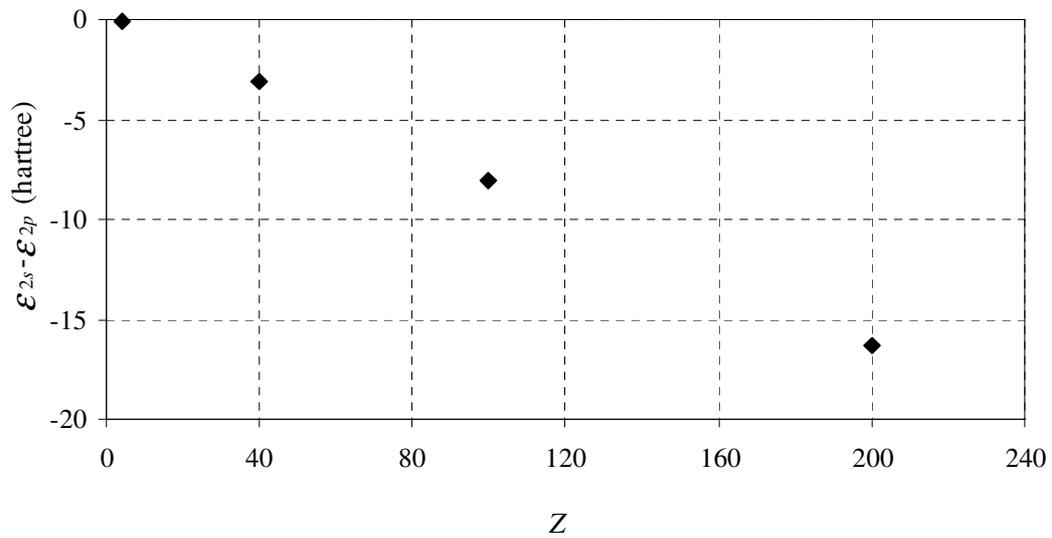

**FIG. 4.** The difference between the occupied $2s$ and the vacant $2p$ energy levels in Be ion series grows proportionally to the atomic number $Z$.

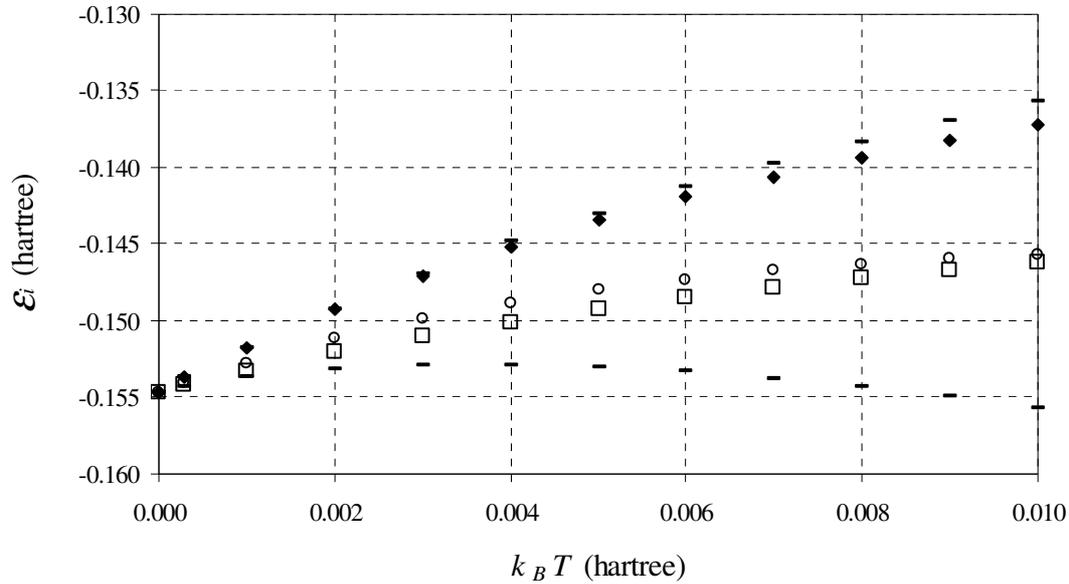

**FIG. 5.** The dependence of the chemical potential (circles), the $3d^{\downarrow}$ (rhombi) and the $4s^{\downarrow}$ (squares) energy levels of Fe on the temperature. The horizontal marks are $1\,k_BT$ above and below the chemical potential.

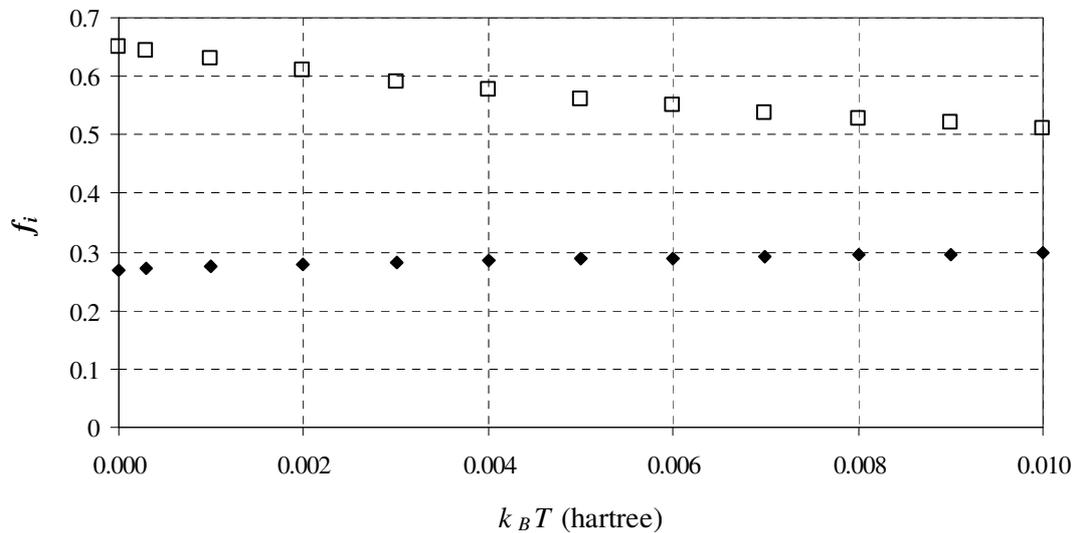

**FIG. 6.** The normalized occupation numbers of $3d^{\downarrow}$ (rhombi) and the $4s^{\downarrow}$ (squares) energy levels of Fe as a function of temperature.